\documentclass[conference]{IEEEtran}
\IEEEoverridecommandlockouts
\usepackage{cite}
\usepackage{amsmath,amssymb,amsfonts}
\usepackage{algorithmic}
\usepackage[T1]{fontenc}
\usepackage{graphicx, url}
\usepackage{textcomp}
\usepackage{xcolor}
\DeclareMathOperator*{\argmin}{arg\,min}
\def\z{{\mathbf z}}
\usepackage{booktabs}
\usepackage{multirow}
\usepackage[hidelinks]{hyperref} 
\def\BibTeX{{\rm B\kern-.05em{\sc i\kern-.025em b}\kern-.08em
    T\kern-.1667em\lower.7ex\hbox{E}\kern-.125emX}}
\begin{document}

\title{Reliable Local Explanations for Machine Listening
\thanks{This work was done when the first author was a PhD student at Queen Mary University of London. The work of EB was supported by RAEng Research Fellowship RF/128. This work was supported by The Alan Turing Institute under the EPSRC grant EP/N510129/1.}
}

\author{\IEEEauthorblockN{Saumitra Mishra\textsuperscript{* \dag}, Emmanouil Benetos\textsuperscript{* \dag}, Bob L. T. Sturm\textsuperscript{\#} \& Simon Dixon\textsuperscript{\dag}}
\IEEEauthorblockA{\textit{\textsuperscript{*}The Alan Turing Institute, London, U.K.} \\
\textit{\textsuperscript{\dag}School of EECS, Queen Mary University of London, U.K.} \\
\textit{\textsuperscript{\#}Divison of Speech, Music and Hearing, KTH Royal Institute of Technology, Stockholm, Sweden} \\ 
smishra@turing.ac.uk, emmanouil.benetos@qmul.ac.uk, bobs@kth.se, s.e.dixon@qmul.ac.uk}
}

\maketitle

\begin{abstract}
One way to analyse the behaviour of machine learning models is through local explanations that highlight input features that maximally influence model predictions. Sensitivity analysis, which involves analysing the effect of input perturbations on model predictions, is one of the methods to generate local explanations. Meaningful input perturbations are essential for generating reliable explanations, but there exists limited work on what such perturbations are and how to perform them. This work investigates these questions in the context of machine listening models that analyse audio. Specifically, we use a state-of-the-art deep singing voice detection (SVD) model to analyse whether explanations from SoundLIME (a local explanation method) are sensitive to how the method perturbs model inputs. The results demonstrate that SoundLIME explanations are sensitive to the content in the occluded input regions. We further propose and demonstrate a novel method for quantitatively identifying suitable content type(s) for reliably occluding inputs of machine listening models. The results for the SVD model suggest that the average magnitude of input mel-spectrogram bins is the most suitable content type for temporal explanations.

\end{abstract}

\begin{IEEEkeywords}
Interpretable Machine Learning, Explainable AI, Machine Listening.
\end{IEEEkeywords}

\section{Introduction}
Recent years have witnessed a remarkable surge in the use of machine learning (ML) models across different application domains (e.g., computer vision, audio). However, due to the ``black-box'' nature of the majority of high performing ML models, there has been a growing concern about using them in safety-critical applications (e.g., healthcare, finance). Motivated by this, researchers have developed several post-hoc methods to analyse the behaviour of pre-trained ML models. Such methods help to validate whether ML models satisfy hard-to-formalise auxiliary properties (e.g., trustworthiness)~\cite{Doshi_arXiv_2017}. The methods also help to identify dataset faults~\cite{Chettri_slt_2018} and help in improving deep neural network (DNN) architectures~\cite{Zeiler_eccv_2014}.

We can group the existing post-hoc model analysis methods into two high-level categories~\cite{Montavon_dsp_2018}. The first category includes methods that aim to analyse the \emph{global behaviour} of ML models. There exist multiple ways to do this. For example, some methods approximate the global behaviour of complex ML models by interpretable (proxy) models (e.g., decision trees)~\cite{Craven_neurips_1995}. Other methods focus on analysing specific ML models. For example, researchers have proposed methods to understand DNNs by analysing their latent representations~\cite{Mahendran_cvpr_2015, Mishra_ismir_2018} or by synthesising examples that maximally (or minimally) activate their components (e.g., neurons, layers)~\cite{Erhan_tr_2009, Mishra_iclr_2019}. The analysis of the global behaviour of ML models provides useful insights, however, there are challenges. For example, in some cases, the insights from the interpretable proxies may be unfaithful due to the oversimplification of the complex ML model. Moreover, the proxy models may themselves become fairly complex (e.g., very deep decision trees) and hinder interpretability~\cite{Gilpin_dsaa_2018}.

The second category includes methods that aim to analyse the \emph{local behaviour} of ML models by explaining their predictions. The explanations generated by such methods highlight influential input features for a prediction. For example, for an image classification DNN, a local explanation method can generate heat maps that highlight pixels positively (or negatively) influencing model predictions~\cite{Sundarrajan_icml_2017, Bach_plos_2015}. Thus, local explanations help gain insights into the behaviour of ML models even with fairly complex global behaviour.

Recently, researchers have demonstrated that for the DNN-based image classification models, the explanations from some local explanations methods do not accurately reflect the underlying model behaviour~\cite{Kindermans_arXiv_2017, Adebayo_neurips_2018}. This suggests that validating the behaviour of local explanation methods is essential to prevent the generation of misleading insights from the local analysis of ML models. Inspired by this research, this work focuses on analysing the behaviour of local explanation methods in the context of machine listening models that automatically process sounds (e.g., recorded music, natural sounds) using computational models to extract meaningful information.

We can group the existing methods to locally analyse machine listening models into two main categories. The first category includes methods that are specific to analysing deep machine listening models. For example, in~\cite{Choi_arXiv_2016}, the authors visualised and auralised saliency maps generated using the deconvolution method~\cite{Zeiler_eccv_2014} to explain predictions of a convolutional neural network (CNN)-based music genre classification model. Similarly, in \cite{Mishra_eusipco_2018}, the authors proposed to invert features from the deepest hidden layer of a DNN to identify input regions (groups of time-frequency bins) maximally influencing model predictions.

This work focuses on the second category of methods that are model-agnostic. Thus, the insights from their analysis will very likely generalise to all types of machine listening models. Specifically, we analyse the behaviour of \emph{SoundLIME (SLIME)} - a model-agnostic method that explains model predictions by highlighting the temporal, spectral or time-frequency components (segments) in inputs that maximally influence model predictions~\cite{Mishra_ismir_2017}. SLIME extends the local interpretable model-agnostic explanations (LIME) algorithm~\cite{Ribeiro_kdd_2016} to machine listening. Earlier works used SLIME to examine the predictions of machine listening models~\cite{Mishra_ismir_2017} and to identify dataset faults~\cite{Chettri_slt_2018}.

SLIME explains a prediction by using a set of synthetic samples that it generates by perturbing an input. These perturbations randomly occlude input components. Previous works performed occlusion by replacing the content in occluded components by the `zero' value~\cite{Mishra_ismir_2017, Chettri_slt_2018}. However, it is not clear if this is appropriate as it implicitly assumes that a machine listening model is insensitive to the zero value, which may not be correct~\cite{Mittelstadt_fat_2019}. Moreover, in~\cite{Mishra_ismir_2017}, the authors discussed a preliminary experiment suggesting that explanations from SLIME are sensitive to the number of synthetic samples. However, due to the use of a small number of inputs from a single dataset, it is not evident if such a behaviour will extend to a large number of inputs and to other datasets.

In this paper, we address the above challenges and make the following contributions.

\begin{itemize}

\item We perform large-scale experiments to analyse the sensitivity of SLIME explanations to changes in two input parameters: the content of the occluded input components and the number of synthetic samples. The results demonstrate that SLIME explanations are sensitive to both parameters, suggesting that a careful selection of suitable values of these parameters is essential to generate reliable explanations. This further suggests that it is highly likely that such a behaviour will extend to other explanation methods that use input occlusion in their explanation pipeline.

\item We introduce and demonstrate a novel method to generate reliable explanations from SLIME. The method quantitatively selects suitable content type(s) by using ground-truth.

\item We introduce five content types for perturbing inputs (by occlusion) of machine listening models.

\end{itemize}

The code for all the experiments is publicly available\footnote{ \url{https://github.com/saum25/local_exp_robustness}}.

\section{SoundLIME}
\begin{figure} [t]
 \centering
 \includegraphics[width=\columnwidth]{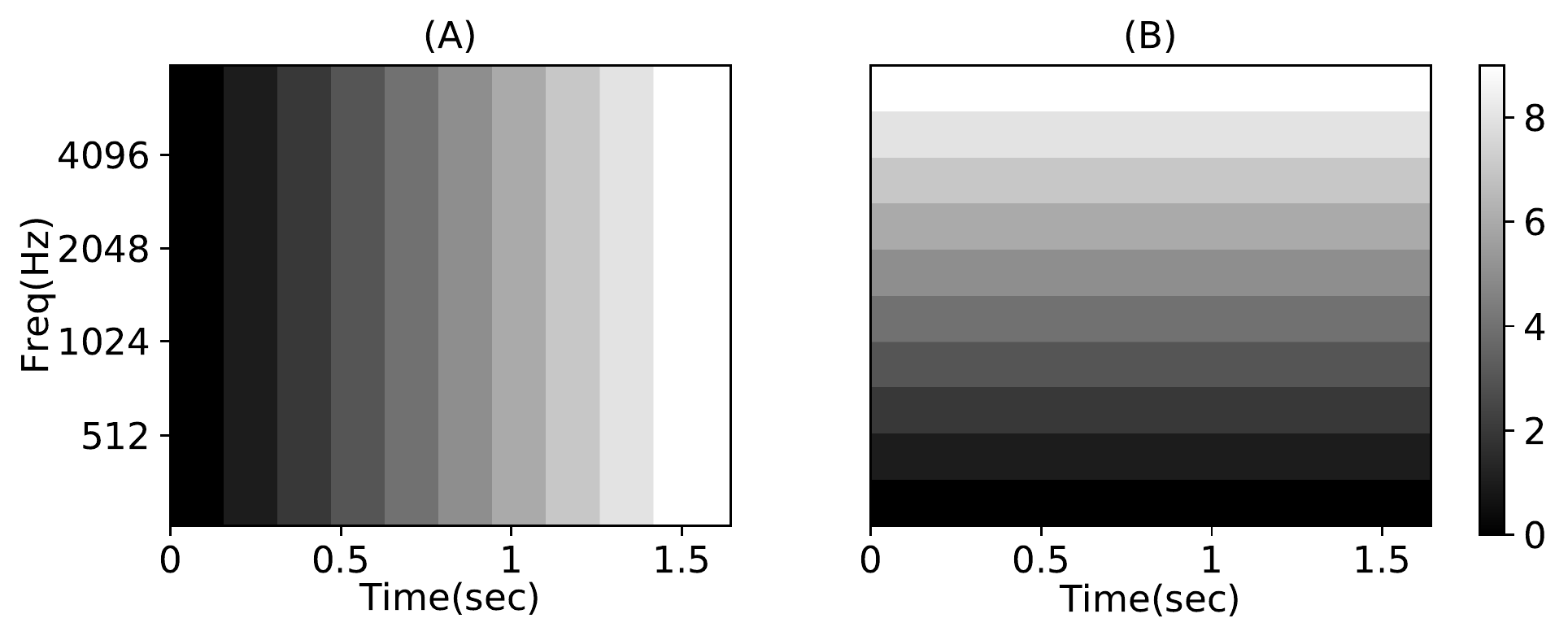}
 \caption{The plots depict how SLIME segments a mel-spectrogram into ten interpretable components. (A) Temporal segmentation and (B) Spectral segmentation. The colour bar depicts the indices of the interpretable components.} 
 \label{fig:slime_segmentation}
\end{figure}

SLIME is a local explanation method that explains the predictions of any machine listening model~\cite{Mishra_ismir_2017}. SLIME generates three types of explanations that identify influential temporal, spectral, and time-frequency content in input audio. SLIME extends the LIME algorithm~\cite{Ribeiro_kdd_2016} to machine listening by defining three \emph{interpretable sequences} (temporal $\mathit{\mathcal{X}^t}$, spectral $\mathit{\mathcal{X}^f}$, and time-frequency $\mathit{\mathcal{X}^{tf}}$) for an input audio $\mathbf{x}\in\mathbb{R}^n$. SLIME generates these sequences by segmenting $\mathbf{x}$ (uniformly or non-uniformly) into a sequence of \emph{interpretable components (ICs)}. For example, Fig.~\ref{fig:slime_segmentation} depicts how SLIME segments a mel-spectrogram\footnote{A mel-spectrogram is a perceptually motivated time-frequency representation of an audio signal~\cite{Logan_ismir_2000}.}, temporally and spectrally into ten ICs (indexed $0-9$). The temporal segmentation in Fig.~\ref{fig:slime_segmentation}(A) generates a sequence of ten temporal segments (each called a \emph{super-sample} and denoted as $T_i$). Similarly, the spectral segmentation in Fig.~\ref{fig:slime_segmentation}(B) generates a sequence of ten spectral segments (each denoted as $A_i$). Thus, $\mathit{\mathcal{X}^t} = (T_0, T_1, ....., T_8, T_9)$ and $\mathit{\mathcal{X}^f} = (A_0, A_1, ....., A_8, A_9)$. SLIME maps each of the three interpretable sequences to the \emph{interpretable space} $\mathcal{T} = \{0, 1\}^{|\mathcal{X}|}$ and generates corresponding \emph{interpretable representations} $\mathbf{x}' \in \mathcal{T}$.

SLIME explains a prediction from a classifier $C:\mathbb{R}^n \rightarrow [0, 1]$ by approximating $C$ by a linear model\footnote{$\gamma$ can be any interpretable model, but in this work $\gamma$ is a linear model.} $\gamma(\mathbf{z}') = \mathbf{w}^T\cdot\mathbf{z}'$ in the interpretable space, where $\mathbf{z}' \in \mathcal{T}$ represents a synthetic sample. SLIME generates $N_s$ synthetic samples by randomly setting the dimensions of $\mathbf{x}'$ to zero. For example, a synthetic sample for the temporal sequence in Fig.~\ref{fig:slime_segmentation}(A) is $\mathbf{z}' = (1, 1, 1, 0, 0, 1, 1, 0, 1, 1)$, where $0$ indicates the absence of super-samples $T_3, T_4$, and $T_7$. There exist $2^{N_c}$ unique synthetic samples for an interpretable sequence with $N_c$ components. The magnitude and polarity of the weights $\mathbf{w}$ of the linear model constitute explanations for a prediction. Formally, SLIME learns $\gamma$ by the optimisation
\begin{equation} \label{eq:1}
\argmin_{\gamma} \; (L(C, \gamma, \rho_{\mathbf{x}}) + \Delta(\gamma))
\end{equation}
where $L(C, \gamma, \rho_{\mathbf{x}})$ is the locally-weighted loss function and and $\Delta(\gamma)$ quantifies the model complexity (e.g., sparsity). SLIME defines the loss function as the weighted squared difference between classifier prediction $C(\mathbf{z})$ and $\gamma(\mathbf{z}')$, where $\mathbf{z}\in\mathbb{R}^n$ represents the perturbed version of an input and is generated by mapping $\mathbf{z}'$ to the input space. $ \rho_{\mathbf{x}}(\z)$ is the distance between a synthetic sample $\z$ and the input $\mathbf{x}$. Formally, SLIME defines the loss function as
\begin{equation}
L(C, \gamma, \rho_{\mathbf{x}}) = \sum_{{(\z^\mathit{\prime}_{i}, \z_{i})}\in Z} \rho_{\mathbf{x}}(\z_{i}) [C(\z_{i}) - \gamma(\z^\mathit{\prime}_{i})]^2
\end{equation} 

\section{Machine listening use case and model}

\begin{figure}[t]
	\centering
	\includegraphics[width=\columnwidth]{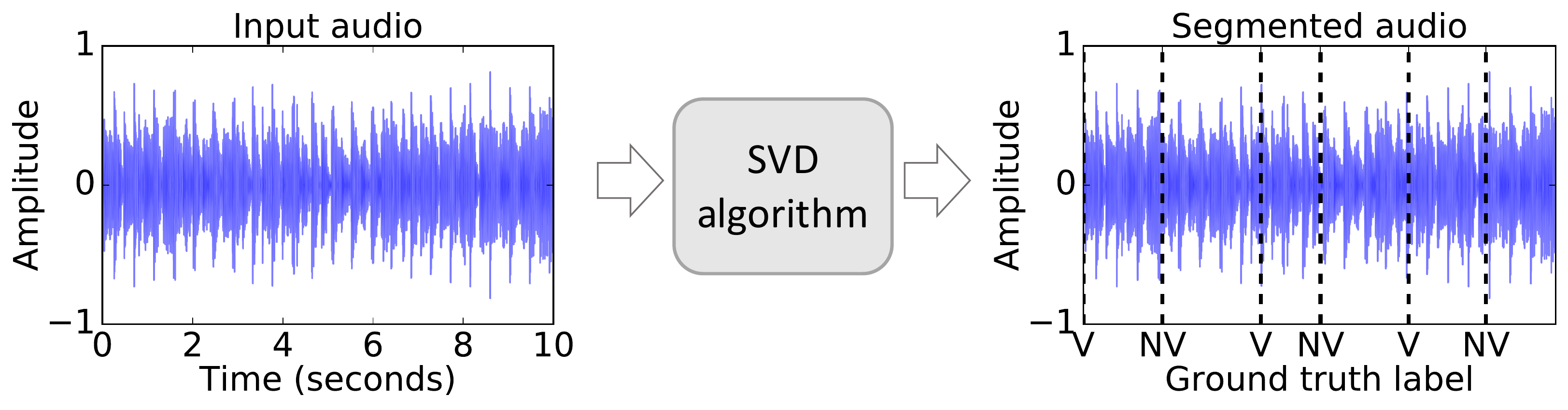}
	\caption{The figure depicts the application of a singing voice detection (SVD) algorithm to a $10$-second musical audio clip. The SVD algorithm segments the input (left) into temporal sections, indicating the presence or absence of singing voice (right). NV and V refer to the labels corresponding to the beginnings of non-vocal and vocal segments, respectively.}
\label{fig:bgd_svd_example}
\end{figure}

We use \emph{singing voice detection (SVD)} as the machine listening use case for the experiments. SVD refers to the automatic detection of the presence (or the absence) of the singing voice (or vocals) in short-duration (e.g., $200$ ms)
audio frames (or excerpts)~\cite{Lehner_talsp_2018}. The application of SVD to musical audio recordings segments them into vocal and non-vocal (instrumental) sections. 
Fig.~\ref{fig:bgd_svd_example} depicts the output from an SVD algorithm for a $10$-second musical audio.

We select the SVD method proposed in~\cite{Schluter_ismir_2015} as the model is open-sourced\footnote{Available at {https://github.com/f0k/ismir2015}} and is the state-of-the-art on publicly available benchmark datasets. In this work, we refer to this model as `SVDNet-R0'. The model is a nine layer deep CNN with an architecture consisting of the convolutional, max-pooling, and fully-connected layers. Each convolutional layer performs convolutions using $3 \times 3$ filters with $1 \times 1$ stride and no zero padding. Each max-pooling layer performs $3 \times 3$ max-pooling with $3 \times 3$ stride and no zero padding. The inputs to the model are normalised log-scaled mel-spectrogram excerpts of around $1.6$-second duration ($115$ frames). A single sigmoidal output neuron provides the probability of the presence of vocals in the central frame of a musical audio excerpt using $57$ frames on either side as context.

The authors trained SVDNet-R0 using mel-spectrogram excerpts and ground-truth labels from the Jamendo training dataset~\cite{Ramona_icassp_2008}. The Jamendo dataset consists of $93$ full-length songs with Creative Commons license collected from the Jamendo free music sharing website\footnote{https://www.jamendo.com}. 
The dataset comes pre-partitioned into subsets of $61$ (training), $16$ (validation), and $16$ (testing) songs, respectively. Each song has sub-second annotations indicating the start and end of the vocal and non-vocal segments. The training methodology minimises the binary cross entropy loss between CNN predictions and ground-truth labels. Additionally, the training uses three data augmentation methods (pitch shifting, time stretching and random filtering) to tackle the problem of the smaller training dataset.

The authors trained SVDNet-R0 using the Theano and Lasagne frameworks, but for this work, we train SVDNet-R0 using the Tensorflow framework. We call the new model `SVDNet-R1'. Table~\ref{tab:svd_res} reports the performance of both the SVD models on the Jamendo test dataset. The performance metrics of the two models differ in the order of $10^{-1}$, suggesting that the models have similar prediction capabilities.

\begin{table}[t]
\caption{Performance of the SVD models on the Jamendo test dataset.}
\footnotesize
\begin{center}
\begin{tabular}{ccc}
  \toprule
    & SVDNet-R0  &  SVDNet-R1\\ 
  \midrule
 Threshold & 0.66 & 0.50\\
 Error & \textbf{8.1\%} & \textbf{8.4\%}\\
 Precision & 0.901 & 0.896\\
 Recall & 0.926 & 0.925\\
 Specificity & 0.912 & 0.908\\
 F-score & 0.913 & 0.910\\
 Framework & Theano \& Lasagne & Tensorflow\\
  \bottomrule
\end{tabular}
\label{tab:svd_res}
\end{center}
\end{table}

\section{Analysing the robustness of SLIME}
\label{sec:slime_rob}

We now analyse whether SLIME explanations for SVDNet-R1 predictions are sensitive to the synthetic sample generation process. SLIME synthesises samples in the interpretable space by randomly occluding ICs in an input. SLIME occludes ICs by replacing them with synthetic components.
We analyse whether SLIME explanations change with changes to the content of the synthetic components.

Earlier work has demonstrated that SLIME explanations are sensitive to the number of synthetic samples $N_s$~\cite{Mishra_ismir_2017}. Thus, we first select an appropriate $N_s$ and then analyse SLIME's behaviour for five types of input perturbations, each modifying the content of the synthetic components. 

\subsection{Selecting an appropriate $N_s$}
\label{ssec:slime_rob_ns}

An experiment in~\cite{Mishra_ismir_2017} demonstrated SLIME's sensitivity to $N_s$ for SVDNet-R0 using a dataset of $80$ randomly selected audio excerpts from the Jamendo test dataset. The results of the experiment suggested that SLIME generated \emph{stable explanations} for $N_s \geq 5000$, where the stability of an explanation is defined to be inversely proportional to the number of unique interpretable components $U_n$ in a set of $k$ explanations, obtained by applying SLIME $k$ times to the same input. $U_n$ computation ignores the order of ICs in an explanation. The experiment used $k=5$.

For the experiments in this paper, we use a different model (SVDNet-R1) and comparatively much larger datasets. Thus, before analysing SLIME's behaviour for different types of input perturbations, it is essential to recompute an appropriate $N_s$ using the new model and datasets. This will ensure that any changes in SLIME explanations while experimenting with different input perturbation methods are not the result of using an inappropriate $N_s$. Moreover, the identification of an appropriate $N_s$ will also help validate if the observation about SLIME's sensitivity to $N_s$ generalises to other datasets.

To select an appropriate $N_s$, we apply SLIME to generate temporal explanations for SVDNet-R1 predictions for $400$ mel-spectrogram excerpts ($25$ randomly selected excerpts from each song in the Jamendo test dataset). SLIME generates ICs by segmenting each excerpt along the temporal axis into $10$ temporal segments (see Fig.~\ref{fig:slime_segmentation}) and occludes ICs by synthetic components with content set to the zero value.
Each explanation highlights the three most influential ICs (we call it the `top-$3$' case) positively or negatively influencing a prediction, where the influence of an IC is determined by the magnitude of the linear model weight associated with it. We compute $U_n$ for each excerpt using an aggregated set of temporal explanations constructed by applying SLIME five times to the same excerpt. We perform the experiment for nine different values of $N_s$ ranging from $1000$ to $70000$.

\begin{figure}[t]
\centerline{\includegraphics[width=\columnwidth]{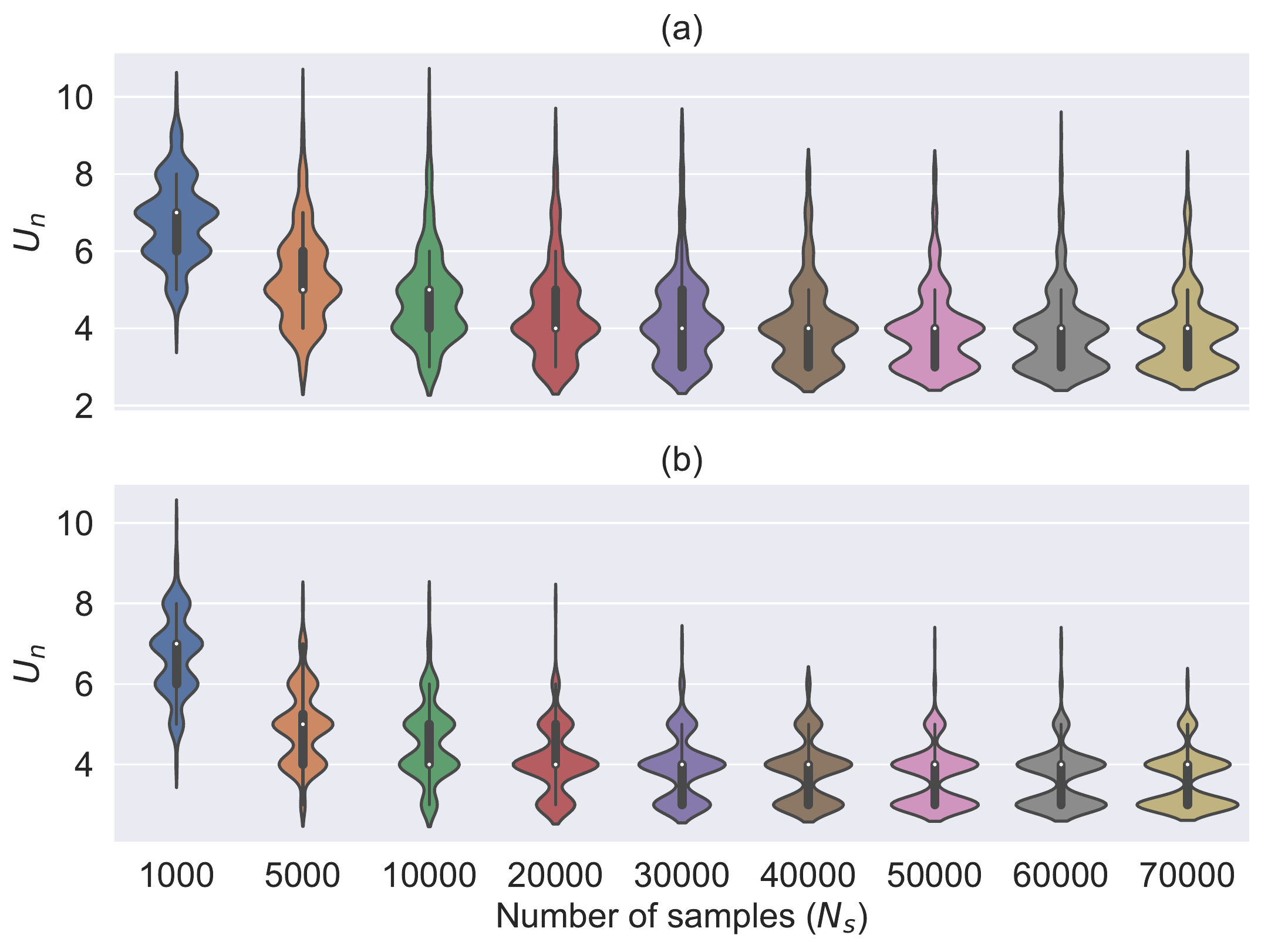}}
 \caption{The violin plots depict the influence of the number of synthetic samples on the stability of SLIME explanations for mel-spectrogram excerpts from the (a) Jamendo and (b) RWC datasets. $U_n$ represents the number of unique interpretable components.}
\label{fig:slime_robust_exp1_n_samp_analysis}
\end{figure}

Fig.~\ref{fig:slime_robust_exp1_n_samp_analysis} (a) presents the results of the experiment. The results demonstrate that similar to the preliminary experiment in~\cite{Mishra_ismir_2017}, SLIME explanations stabilise (have lower median $U_n$) for higher $N_s$ values. Moreover, for the new model and dataset, appropriate values of $N_s$ seem to be $\geq 50000$, as for those values, the $U_n$ distribution has the median value of four and a high likelihood of $U_n = 3$.

We repeat the above experiment for excerpts from the RWC popular music dataset~\cite{Mauch_ismir_2011} to analyse if the above observations extend to other datasets. The RWC dataset contains $100$ full-length popular
music songs out of which $80$ songs are in Japanese and $20$ songs are in English. The dataset is not available pre-partitioned into separate subsets. Thus, we create a test set (we call this `RWC-test') for experiments in this paper by randomly sampling $20$ songs from the dataset. For the experiment in this section, we randomly sample $500$ excerpts from RWC-test ($25$ excerpts from each song in RWC-test). Fig.~\ref{fig:slime_robust_exp1_n_samp_analysis}(b) presents the results of the experiment. Similar to Jamendo, $N_s$~$\geq 50000$ seems to generate stable explanations for the RWC dataset.

\subsection{Analysing explanation sensitivity to synthetic content}
\label{ssec:slime_rob_ana}

\begin{figure*}
\centerline{\includegraphics[width=\textwidth]{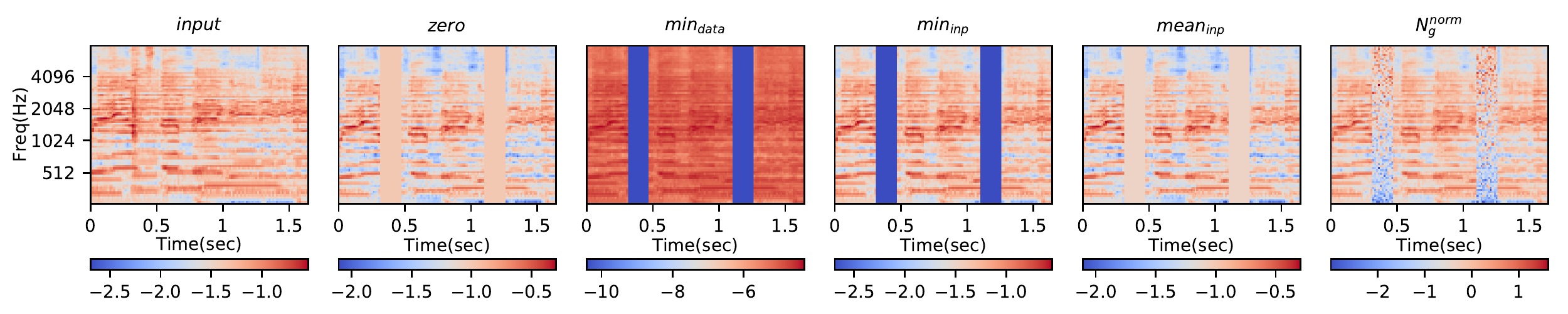}}
 \caption{Visualisations depicting an input mel-spectrogram from the Jamendo test dataset and its five perturbed versions, each generated by SLIME by occluding input temporal components with indices $2$ and $7$ by synthetic components filled with five different content types.}
\label{fig:slime_content_types}
\end{figure*}

We now analyse if SLIME explanations are sensitive to the content of synthetic components. For example, we examine whether SLIME explanations change if instead of the zero value, SLIME uses Gaussian noise to occlude ICs? Such an analysis will help understand whether the selection of the content of synthetic components is critical for SLIME to generate reliable explanations.

SLIME aims to utilise the effect of the removal of randomly selected ICs on model predictions. However, as machine listening models predict using fixed-size inputs, SLIME approximates the ``removal effect'' by filling the selected ICs with synthetic content, hypothesising that the occluded ICs have minimal influence on model predictions~\cite{Fong_iccv_2017}.

In this section, we examine whether the above hypothesis depends on the synthetic content type. To do that, we propose five different types of synthetic content
and group them into two categories: (1) content that generates very low energy (near silent) components and (2) content that generates components with non-discriminatory audio features. The first category includes three types of input content: 
\begin{enumerate}
    \item the $\mathit{zero}$ value
    \item the minimum bin magnitude across a dataset ($\mathit{min_{data}}$)
    \item the minimum bin magnitude in an input ($\mathit{min_{inp}}$).
\end{enumerate} 
It is conceivable that audio feature extraction from the above very low energy components will result in insignificant features with minimal influence on model predictions. The second category includes two types of input content: 
\begin{enumerate}
    \item the average bin value in an input ($\mathit{mean_{inp}}$)
    \item zero-mean unit-variance Gaussian noise ($N_g$).
\end{enumerate}
These contents will generate ICs lacking dominant structures and feature extraction from such ICs will generate non-discriminatory audio features with limited influence on model predictions. Fig.~\ref{fig:slime_content_types} depicts a mel-spectrogram excerpt from the Jamendo dataset and its five perturbed versions generated by occluding the temporal interpretable components with indices $2$ and $7$ with the proposed content types. It is important to note that in this work, we use the standardised version of $N_g$, we call it $N^{norm}_g$. This involves normalising each frequency bin in $N_g$ using the mean and standard deviation computed across each frequency band over the Jamendo training dataset.

\begin{table*}
\caption{SLIME explanations for randomly selected mel-spectrogram excerpts from the Jamendo and RWC datasets for five content types. Index: instance index; $\mathit{V_{prob}}$: model confidence about the presence of singing voice; $\mathit{E_{type}}$: explanation type; Explanations: top-$3$ interpretable components maximally influencing a prediction; and $\mathit{zero}$, $\mathit{min_{data}}$, $\mathit{min_{inp}}$, $\mathit{mean_{inp}}$, and $\mathit{N^{norm}_g}$ refer to the content types that occlude an input by using the zero value, minimum bin magnitude across a dataset, minimum bin magnitude in an input, average bin magnitude in an input and, standardised zero-mean unit-variance Gaussian noise, respectively.}
 \begin{center}
 \begin{tabular}{c c c c c c c c c c}
  \toprule
  \multirow {2}{*}{Dataset} &  \multirow{2}{*}{Audio file} & \multirow {2}{*} {Index} & \multirow {2}{*}{$\mathit{V_{prob}}$} & \multirow{2}{*}{$\mathit{E_{type}}$} & \multicolumn {5} {c} {Explanations}  \\ [0.5ex]
  \cline{6-10} \\ [-2.0 ex]
  & & & & & $\mathit{zero}$ & $\mathit{min_{data}}$ & $\mathit{min_{inp}}$ & $\mathit{mean_{inp}}$ & $\mathit{N^{norm}_g}$ \\
  \midrule
  Jamendo & $03$ - School.mp3 & $435$ & $0.023$ & temporal & $4, 3, 6$ & $4, 1, 5$ & $1, 3, 5$ & $6, 9, 1$ & $1, 0, 9$ \\
  Jamendo & $03$ - School.mp3 & $3162$ & $0.915$ & temporal & $1, 4, 3$ & $4, 7, 2$ & $5, 4, 2$ & $5, 7, 8$ & $1, 8, 2$ \\
  \midrule
  Jamendo & $03$ - Une charonge.ogg & $19291$ & $0.017$ & spectral & $4, 6, 8$ & $4, 5, 7$ & $5, 1, 4 $&$ 6, 8, 3$& $8, 5, 1$ \\
  Jamendo & $03$ - Une charonge.ogg & $1888$ & $0.861$ & spectral & $2, 4, 8$ & $6, 2, 3$ & $7, 2, 6$ & $2, 6, 9$ & $7, 6, 1$\\
  \midrule
  RWC & RWC-MDB-P-2001-M01/016 Audio Track.aiff & $4732$ & $0.233$ & temporal & $1, 5, 0$ & $2, 5, 4$ & $4, 2, 7$ & $1, 0, 8$ & $6, 5, 4$ \\
  RWC & RWC-MDB-P-2001-M01/016 Audio Track.aiff & $701$ & $0.871$ & temporal & $5, 1, 8$ & $2, 5, 3$ & $5, 2, 4$ & $1, 2, 0$ & $6, 7, 9$ \\
  \midrule
  RWC & RWC-MDB-P-2001-M04/4 Audio Track.aiff & $1578$ & $0.019$ & spectral & $7, 8, 9$ & $1, 2, 4$ & $5, 1, 0$ & $7, 8, 9$ & $1, 4, 2$ \\
  RWC & RWC-MDB-P-2001-M04/4 Audio Track.aiff & $12794$ & $0.966$ & spectral & $1, 7, 8$ & $3, 2, 1$ & $1, 2, 6$ & $1, 7, 8$ & $5, 7, 1$ \\
  \bottomrule
 \end{tabular}
\label{tab:slime_rob_tab1}
\end{center}
\end{table*}

We apply SLIME to explain SVDNet-R1 predictions by performing input occlusion using the five content types from above. We analyse if the modification of the content of synthetic components changes SLIME predictions. We generate top-$3$ explanations for eight randomly selected mel-spectrogram excerpts from four audio files in the Jamendo and RWC test datasets. We sample a vocal and a non-vocal instance from each audio file. SLIME generates temporal explanations for two excerpts from each dataset and generates spectral explanations for the remaining four excerpts. SLIME uses $N_s = 70000$ as it generates stable explanations for the new model (see Section \ref{ssec:slime_rob_ns}). SLIME generates the temporal and spectral explanations by segmenting an input into ten ICs along the time and frequency axes, respectively (see Fig.~\ref{fig:slime_segmentation}). All other parameters for executing SLIME are same as the ones in Section \ref{ssec:slime_rob_ns}. 

Table~\ref{tab:slime_rob_tab1} reports the results of the experiment. Each explanation presents the top-3 ICs arranged in the decreasing order of their influence on the prediction. The results suggest that for all the eight instances, there is little overlap in explanations generated using different content types. For example, for each of the first six instances, there are no common ICs among the explanations generated for the five content types. The explanations for those instances do include some ICs that are more frequent, but their influence in a prediction (their occurrence order) keeps changing. For example, for the instance with index $1888$, the IC with index $6$ is present in four explanations, but there is a lack of consistency about how much the component influences the prediction. 
The spectral explanations for RWC are comparatively more coherent. For example, for the instance with index $1578$, all the three ICs and their ordering match between the explanations generated using the $\mathit{zero}$ and $\mathit{mean_{inp}}$ content types.
However, for this instance, such behaviour does not generalise to explanations for the other content types.
Similarly, the instance with index $12794$ has the IC with index $1$ common among all the explanations, but with varying influence in the prediction. These results suggest that for the selected excerpts SLIME explanations are sensitive to the content of synthetic components.

It is important to note that the sensitivity of SLIME explanations to the content of synthetic components in the above experiment may still be the result of an inappropriate $N_s$ as the experiment in Section \ref{ssec:slime_rob_ns} identified appropriate $N_s$ values only for temporal explanations generated using the zero value as the content type. We verify this by analysing whether $N_s = 70000$ is also an appropriate value for the other content types and spectral explanations across both the datasets. Thus, for an excerpt sampled randomly from one of the test datasets and a content type, we apply SLIME to first generate five temporal and five spectral explanations and then calculate $U_n$ for each explanation type by aggregating the corresponding explanations. To limit the computational time, we execute the above steps for two randomly selected excerpts from each song in the Jamendo and RWC test datasets. Thus, we analyse the explanation stability for $32$ and $40$ randomly selected instances from the Jamendo and RWC test datasets, respectively.

\begin{figure} [t]
 \centering
 \includegraphics[width=\columnwidth]{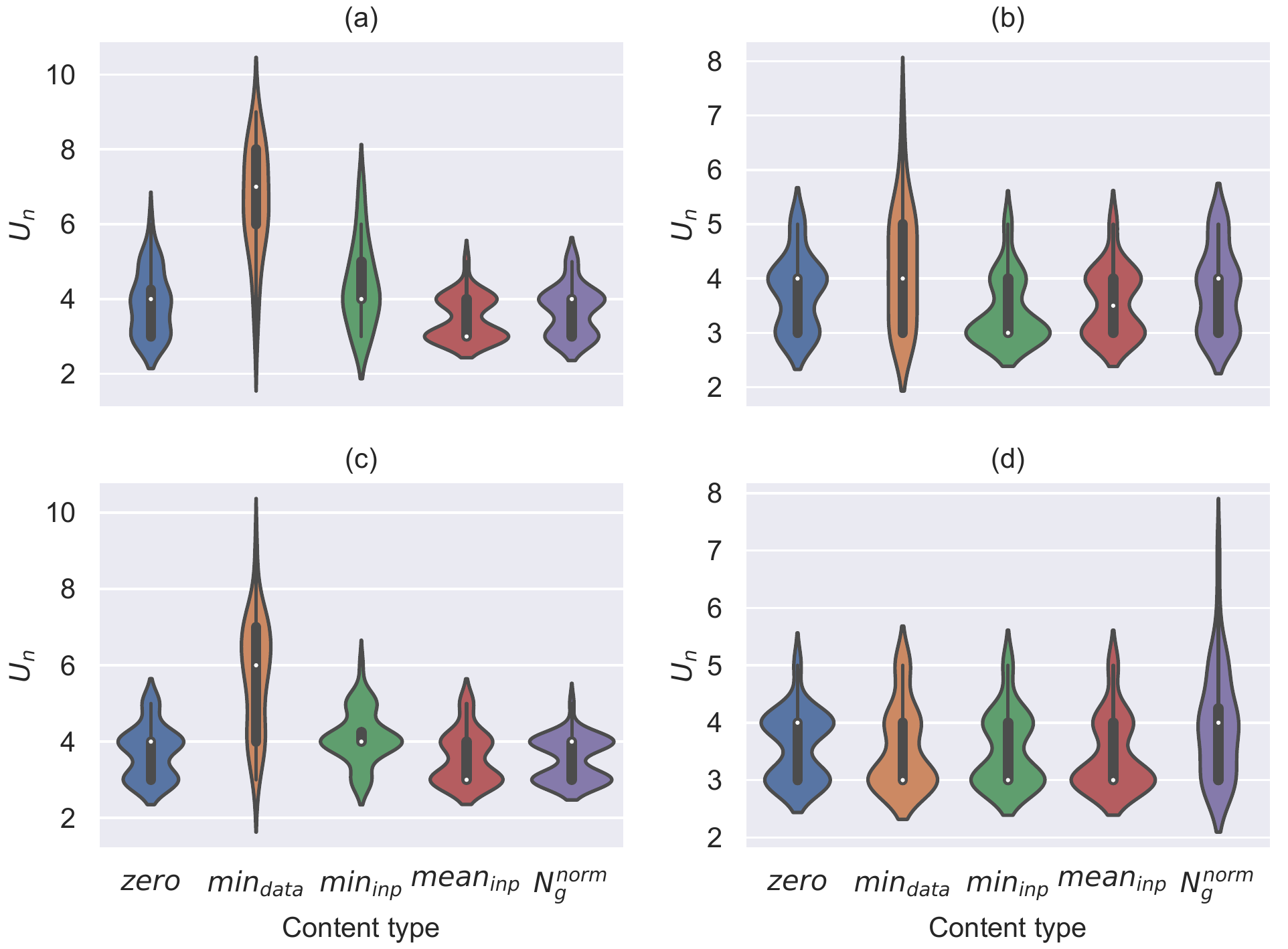}
 \caption{The violin plots depict the influence of content type on the stability of SLIME explanations for four cases. (a) and (c) depict results for temporal explanations from Jamendo and RWC, respectively. (b) and (d) depict results for spectral explanations from Jamendo and RWC, respectively. $U_n$ represents the number of unique interpretable components.}
 \label{fig:slime_fv_analysis}
\end{figure}

Fig.~\ref{fig:slime_fv_analysis} depicts the results of the experiment. Fig.~\ref{fig:slime_fv_analysis}(a) and (b) depict the $U_n$ distribution for the temporal and spectral explanations, respectively from Jamendo. Similarly, Fig.~\ref{fig:slime_fv_analysis}(c) and (d) depict the $U_n$ distribution for the temporal and spectral explanations, respectively from RWC. The results demonstrate that except $\mathit{min_{data}}$, the explanations for the other content types across both the datasets have $U_n$ distributions similar to those for the zero content type. For example, Fig.~\ref{fig:slime_fv_analysis}(c) suggests that for temporal explanations from RWC, the most likely values of $U_n$ are $4$ or $3$, suggesting that except for $\mathit{min_{data}}$, $N_s = 70000$ is also an appropriate value for the other content types. Interestingly, the explanations for $\mathit{min_{data}}$ seem unstable (have high $U_n$) for all except the spectral explanations for RWC excerpts (Fig.~\ref{fig:slime_fv_analysis}(d)). These results suggest that $N_s = 70000$ is not an appropriate value for $\mathit{min_{data}}$ and this may have contributed to the sensitivity of explanations for this content type in Table~\ref{tab:slime_rob_tab1}. The experiment also demonstrates that for all the other content types, $N_s = 70000$ is an appropriate value and thus does not contribute to the explanation sensitivity for those content types.

\begin{figure} [t]
 \centering
 \includegraphics[width=0.95\columnwidth]{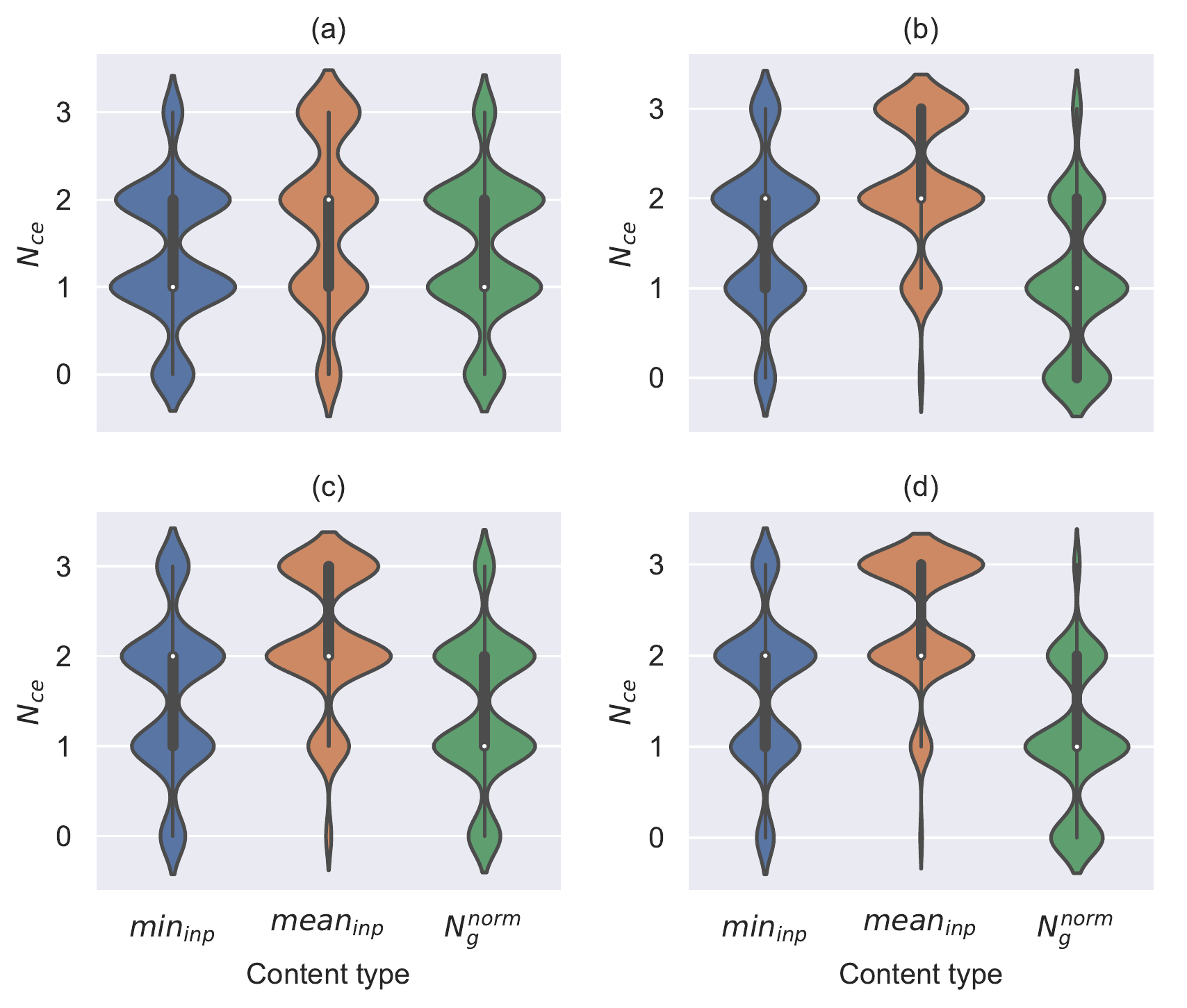}
 \caption{The violin plots depict the influence of content type on SLIME explanations for four cases. (a) and (c) depict results for temporal explanations for Jamendo and RWC, respectively and (b) and (d) depict results for spectral explanations for Jamendo and RWC, respectively. $\mathit{N_{ce}}$ refers to the number of common interpretable components between the explanation with content type zero and the explanation with content type given on the horizontal axis.}
 \label{fig:slime_fv_exps_its}
\end{figure}

We further analyse the effect of different content types on SLIME explanations on a large scale to verify whether the conclusions about the sensitivity of SLIME explanations to the content type generalise. To do that, we first randomly sample $50$ excerpts from each song in each test dataset, generating subsets with $800$ and $1000$ excerpts corresponding to Jamendo and RWC, respectively. We then generate temporal and spectral explanations for the SVDNet-R1 predictions for excerpts in each subset. Each explanation highlights the top-$3$ ICs (positive or negative) maximally influencing a prediction. We generate four explanations for each excerpt, where each explanation corresponds to one of the four content types ($\mathit{zero}$, $\mathit{min_{inp}}$, $\mathit{mean_{inp}}$, and $\mathit{N^{norm}_g}$) that generated stable explanations for $N_s = 70000$. 

Fig.~\ref{fig:slime_fv_exps_its} depicts the results of the experiment. The subplots (a) and (b) correspond to the temporal and spectral explanations, respectively, for the Jamendo excerpts and subplots (c) and (d) correspond to the temporal and spectral explanations, respectively, for the RWC excerpts. Each plot depicts the distribution of the number of common interpretable components $\mathit{N_{ce}}$ between explanations generated using two different content types, one of which is the reference content type. We use the $\mathit{zero}$ content type as the reference. Thus, for each explanation, we compute how many ICs remain the same when SLIME replaces the reference content type with the other content types ($\mathit{min_{inp}}$, $\mathit{mean_{inp}}$, and $\mathit{N^{norm}_g}$).

The analysis of the results provides interesting insights into the behaviour of SLIME. For example, the explanations generated using $\mathit{mean_{inp}}$ are closer to those generated using the reference content type with low likelihood of no overlap in explanations, although for temporal explanations for the Jamendo excerpts, there exists a fair likelihood of $\mathit{N_{ce}} =  1$. On the other hand, the explanations generated using the other content types have comparatively lower overlap with the explanations using the reference content type. Moreover, for the content type $\mathit{N^{norm}_g}$, the likelihood of no overlap in explanations is fairly high. Overall, the results demonstrate that for both the explanation types and datasets, SLIME explanations are sensitive to the content of synthetic components. This suggests that selecting suitable content type(s) is critical for generating reliable explanations from SLIME.

\section{Generating reliable explanations}
\label{sec:slime_rob_sol}

We now propose a novel method for identifying suitable content type(s) for generating local explanations from SLIME. The method involves two steps.
Step $1$ uses domain knowledge to select a list of `relevant' content types, as a content type relevant to one domain or explanation type may not be relevant to the other. For example, the zero value may be a relevant content type for an RGB image with pixel values between $0$ (lowest intensity) to $255$ (highest intensity), but may not be relevant for a log-scaled mel-spectrogram with values between -min to +max, as it may make quieter regions louder.

Step 2 involves using the ground-truth annotations for selecting suitable content type(s) from the list of relevant content types. For example, for SVD, ground-truth indicates temporal segments containing vocals. SLIME can use this information to select suitable content type(s) for temporal explanations by first generating temporal explanations using all relevant content types and then selecting content type(s) that generate temporal explanations having high overlap with ground-truth. For example, say for an input excerpt with five temporal segments, the ground-truth annotations denote that temporal segments $2$ and $4$ contain vocals and the other temporal segments contain non-vocals. Further assume that an SVD model is $95\%$ confident that the input excerpt contains vocals. The temporal explanations from SLIME highlighting the top-$2$ super-samples positively influencing the prediction\footnote{The super-samples are the output of input segmentation at the ground-truth boundaries.} for three relevant content types are $\{1, 2\}$, $\{2, 4\}$, and $\{3, 5\}$. This suggests that the content type two is the most suitable as its explanations completely overlap with ground-truth.

We now demonstrate the proposed method for temporal explanations for the SVDNet-R1 predictions. However, instead of using audio excerpts and the ground-truth annotations from the Jamendo and RWC datasets, we synthesise audio excerpts and the corresponding ground-truth annotations. This approach has two main benefits. First, it allows to synthesise a large dataset for experimentation which is not possible with the existing SVD datasets. This is due to the short duration of the SVD model inputs (around $1.6$ secs), causing the majority of input excerpts to contain either vocals or non-vocals and thus making them not useful for experimentation. Second, the ground-truth annotations for the Jamendo and RWC datasets may be noisy and thus, using a synthesised dataset helps perform controlled experimentation.

We generate a synthetic dataset using the ccMixter dataset~\cite{Liutkus_tsp_2014} that includes a vocal and a non-vocal stem for each of the $50$ songs it contains. The vocal and non-vocal stems contain all the vocal and non-vocal (instrumental) sounds in a song, respectively. We synthesise the dataset by first randomly selecting ten mel-spectrogram excerpts from each stem file corresponding to a song.
We select excerpts from the same temporal locations in both the stem files. 
Thus, we sample ten pairs of excerpts, where each pair contains excerpts belonging to the vocal and non-vocal stems and sampled at the same time index. We segment each excerpt in a pair into ten temporal segments. We randomly select three temporal segments from the vocal stem and use their content to replace the content of corresponding temporal segments in the non-vocal stem. We repeat this process four times and generate four excerpts, where each excerpt contains seven non-vocal segments and three vocal segments. Thus, for each song in the dataset, we generate $40$ mel-spectrograms and executing the method on the whole dataset generates $2000$ mel-spectrograms and their annotations that indicate the temporal segments containing vocals. We aim to analyse only the true positive excerpts, i.e., those excerpts that contain vocals and are correctly predicted by the SVDNet-R1 model. Thus, the final dataset we use for selecting suitable content type(s) has $656$ excerpts.

\begin{figure}[t]
 \centering
 \includegraphics[width=0.94\columnwidth]{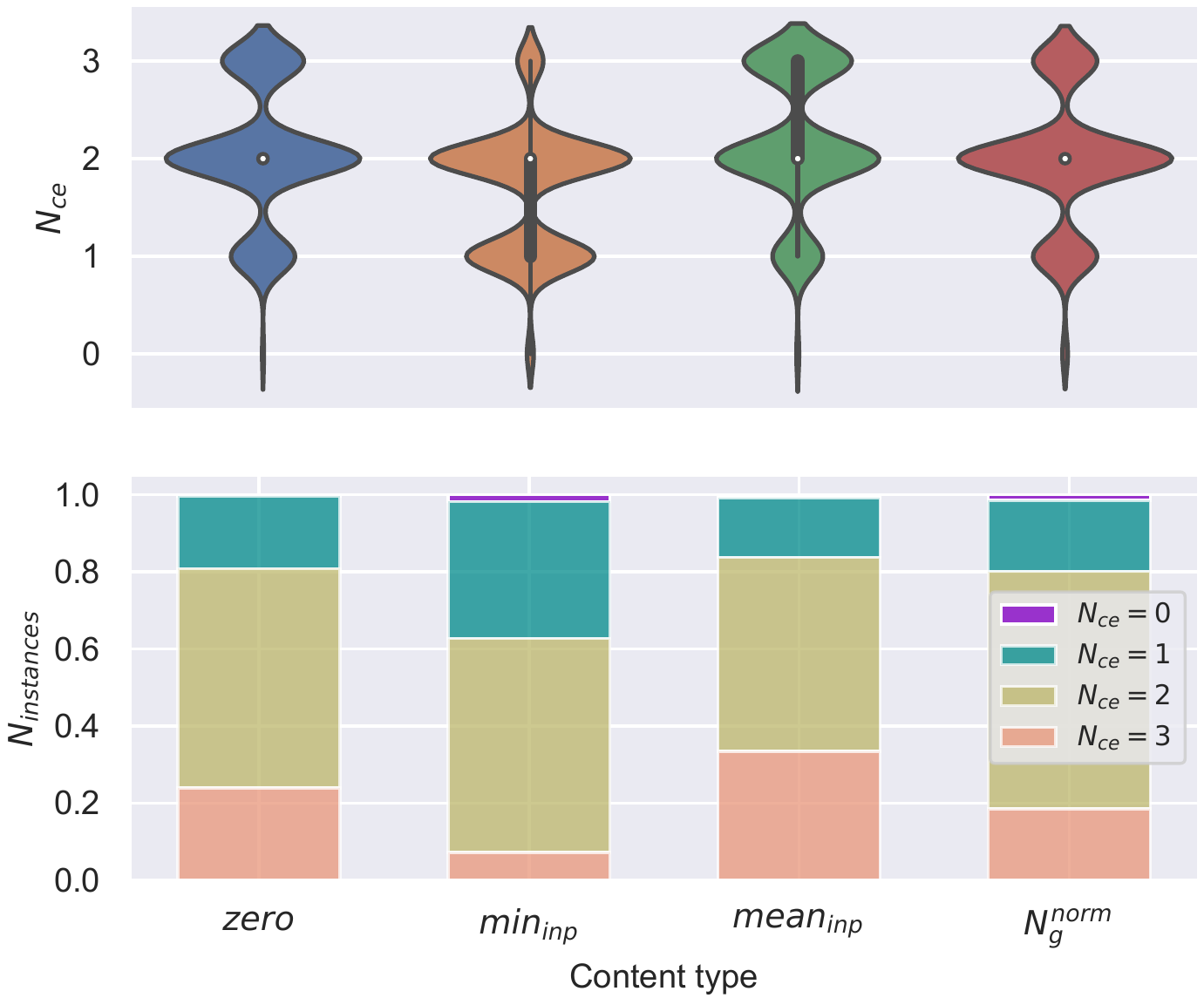}
 \caption{The plots depict the influence of the content types on temporal explanations from SLIME for instances from the synthetic dataset. The top plot depicts the distribution of the number of common interpretable components $\mathit{N_{ce}}$ between the ground-truth explanation and the temporal explanation generated with the content type given on the horizontal axis. The bottom plot depicts the proportion of audio excerpts for different $N_{ce}$ values corresponding to all the four content types.}
 \label{fig:slime_rob_optimal}
\end{figure}

We use SLIME to generate the temporal explanations for the prediction of each excerpt in the synthetic dataset. The temporal explanations highlight the top-$3$ ICs positively influencing SVDNet-R1 predictions. SLIME generates temporal explanations for each of the four content types ($\mathit{zero}$, $\mathit{min_{inp}}$, $\mathit{mean_{inp}}$, and $\mathit{N^{norm}_g}$) that resulted in stable explanations for $N_s = 70000$ (see Section~\ref{ssec:slime_rob_ana}). 
The proposed method selects suitable content type(s) by computing $\mathit{N_{ce}}$ between the ground-truth-based explanations and the temporal explanations for each content type. Fig.~\ref{fig:slime_rob_optimal} presents the results of the experiment. Fig.~\ref{fig:slime_rob_optimal} (top) presents $\mathit{N_{ce}}$ distributions corresponding to each content type. Fig.~\ref{fig:slime_rob_optimal} (bottom) presents the proportion of excerpts for each $N_{ce}$ corresponding to each content type.

The results provide useful insights into the behaviour of SLIME for the different content types and assists in selecting suitable content type(s). The results demonstrate that for around $34\%$ of the excerpts, the temporal explanations corresponding to $\mathit{mean_{inp}}$ completely match the ground-truth-based explanations\footnote{The indices of the ICs in the explanations match, but their occurrence order may differ.}. For the content types $\mathit{zero}$, $\mathit{min_{inp}}$, and $\mathit{N^{norm}_g}$, this number is $23.9\%$, $7.16\%$, and $18.44\%$, respectively. This suggests that among all the content types, $\mathit{mean_{inp}}$ generates the most accurate temporal explanations. The accuracy of SLIME seems low even for the most suitable content type, but it is important to note that SLIME explanations depend on model predictions to the perturbed versions of an input. Thus, less accurate model predictions will result in less accurate SLIME explanations. In this experiment, the predictions from~SVDNet-R1 will be less accurate for the synthetic excerpts as we do not train~SVDNet-R1 on ccMixter, and thus, the synthetic samples are out-of-distribution samples. An example of the difference in the training and test distributions is the composition of the vocal category. In the Jamendo dataset we use to train~SVDNet-R1, the vocal category contains a mix of the vocal and non-vocal sounds, however, in the synthetic dataset, the vocal category contains only vocals. Thus, using the same data distribution for training and testing may result in more accurate SLIME explanations. 

The results also suggest that the likelihood of matching two out of the three ICs in the temporal explanations from SLIME is high for all the content types. Specifically, the proportion of instances whose temporal explanations have at least two ICs in common with the ground-truth-based explanations is $80.79\%$, $62.65\%$, $83.68\%$, and $80.03\%$ for the content types $\mathit{zero}$, $\mathit{min_{inp}}$, $\mathit{mean_{inp}}$, and $\mathit{N^{norm}_g}$ respectively. This suggests that the content types $\mathit{zero}$, $\mathit{mean_{inp}}$, and $\mathit{N^{norm}_g}$ perform equally in reliably identifying at least two ICs positively influencing a prediction. Moreover, the results for these content types also suggest that SLIME accurately identifies at least two influential ICs even though the inputs are out-of-distribution samples. The extremely low likelihood of no overlap between the ground-truth-based explanations and the temporal explanations generated using the different content types further supports the previous statement about the accuracy of SLIME.

The results of the experiment suggest that the content type $\mathit{mean_{inp}}$ is the most suitable content type in generating temporal explanations that perfectly match the ground-truth. This suggests that SVDNet-R1 is comparatively more sensitive to the other content types \cite{Mittelstadt_fat_2019}, and hence, the hypothesis that occluding ICs with those content types is equivalent to removing the corresponding input segments, seems weak. Moreover, the results also suggest that except for $\mathit{min_{inp}}$, all the other content types accurately identify two influential ICs. 

\section{Conclusions and future work}
\label{sec:slime_summary}

In this work, we analysed the behaviour of the SLIME method for explaining the predictions of machine listening models. SLIME perturbs inputs by randomly replacing their ICs with synthetic components. We analysed if SLIME explanations are sensitive to the content of synthetic components. To do this, we applied SLIME to generate temporal and spectral explanations for the predictions of a state-of-the-art deep SVD model. We randomly selected instances from two publicly available benchmark datasets and generated explanations corresponding to four appropriate content types. The results demonstrated that SLIME explanations are sensitive to the content type. This further suggests that it is highly likely for any local explanation method using input occlusion in its explanation generation pipeline to exhibit such a behaviour (e.g.,~\cite{Ribeiro_kdd_2016, Zeiler_eccv_2014}). We also validated that SLIME explanations are sensitive to the number of synthetic samples. These results suggest that careful selection of $N_s$ and the content type is essential for generating reliable SLIME explanations.

We further proposed a method that uses the ground-truth annotations for selecting suitable content type(s) for SLIME. We demonstrated the method for temporal explanations for the deep SVD model predictions. The results suggested that the average bin magnitude of an excerpt is the most appropriate content type as the temporal explanations corresponding to it have at least two ICs matching with ground-truth-based explanations for around $84\%$ of excerpts.

Future work will extend this work by investigating whether other local explanation methods that perturb the input by occlusion~\cite{Zeiler_eccv_2014, Ribeiro_kdd_2016} also exhibit this sensitivity to the content type. Moreover, we plan to examine the behaviour of SLIME for inputs transformed through label-preserving transformations (e.g., increasing the loudness of an input).

\bibliographystyle {IEEEtran}
\bibliography {Refer_short}

\end{document}